\documentstyle[12pt]{article}
\newlength{\extraspace}
\newlength{\extraspaces}
\newcommand{\be}{\begin{equation}
\addtolength{\abovedisplayskip}{\extraspaces}
\addtolength{\belowdisplayskip}{\extraspaces}
\addtolength{\abovedisplayshortskip}{\extraspace}
\addtolength{\belowdisplayshortskip}{\extraspace}}
\newcommand{\ee}{\end{equation}}

\newcommand{\ba}{\begin{eqnarray}
\addtolength{\abovedisplayskip}{\extraspaces}
\addtolength{\belowdisplayskip}{\extraspaces}
\addtolength{\abovedisplayshortskip}{\extraspace}
\addtolength{\belowdisplayshortskip}{\extraspace}}
\newcommand{\ea}{\end{eqnarray}}

\newcommand{\nonu}{\nonumber \\[.5mm]}
\newcommand{\A}{&\!\!\!}

\newcommand{\newsection}[1]{
\vspace{7mm} \pagebreak[3] \addtocounter{section}{1}
\setcounter{subsection}{0} \setcounter{footnote}{0}
\begin{center}
{\large {\bf \thesection. #1}}
\end{center}
\nopagebreak
\medskip
\nopagebreak \hspace{3mm}}

\begin{document}
\begin{center}
{{\bf  Regularization  of Kerr-NUT spacetimes and their thermodynamical quantities }}
\end{center}
\centerline{Gamal G.L. Nashed}
\bigskip

\centerline{\it Center for Theoretical Physics, British University of  Egypt}
 \centerline{Sherouk City 11837, P.O. Box 43, Egypt\footnote{ Mathematics Department,
Faculty of Science, Ain
Shams University, Cairo, 11566, Egypt \vspace*{0.2cm} \\
 Egyptian Relativity Group (ERG) URL:
http://www.erg.eg.net}}

\bigskip
\centerline{ e-mail: nashed@bue.edu.eg}

\hspace{2cm} \hspace{2cm}
\\
\\
\\
\\
\\
\\
\\
\\
\\

In the context of the teleparallel equivalent of  general
relativity (TEGR) theory, continues calculations of  the total energy and
momentum for Kerr-NUT spacetimes  using three different  methods, the gravitational energy-momentum,
 the Riemannian connection  1-form,
${\widetilde{\Gamma}_\alpha}^\beta$ and the Euclidean continuation method,  have been achieved.
 Many local Lorentz transformations, that play the role
of  regularizing tool,  are given  to get the commonly known form of energy and momentum.  We calculate
the thermodynamic quantities  of Kerr-NUT spacetime.
 We also investigate the
first law of thermodynamics and quantum statistical relation.

\newsection{Introduction}

 For several decades, four dimensional solutions of  Einstein field equations have been widely  inquiry  in
gravity community.  Taub-NUT metric$^{ \rm { 1)}}$ is an analytic solution of the vacuum Einstein equations. When the metric is expressed in
Schwarzschild-like coordinates, one has coordinate singularity that occurs at certain values of  radial coordinate
where $g_{rr}$ component   becomes  infinity and corresponding to bifurcate Killing horizons. The Taub-NUT spacetime
is participate in many recent studies of general relativity (GR). Hawking has suggested  that the Euclidean geometry of Taub-
NUT metric could lead to  gravitational analogue of the Yang-Mills instanton.$^{ \rm { 2)}}$ In that case the Einstein
 equations are met with zero cosmological constant and the manifold is $R_4$ with the limit  that  is a twisted three-sphere
  possessing a distorted metric. The Kaluza-Klein monopole has been obtained by embedding the Taub-NUT gravitational instanton
   into five dimensional Kaluza-Klein theory.$^{ \rm { 3,4)}}$ The non-Abelian goals space duals of the Taub-NUT spacetime have
   been also examined  in terms of the local isometry group $SU(2) \times U(1)$.$^{ \rm { 5)}}$ The Taub-NUT spacetime has been shown to be
   associated with $SU(2)$ through T-duality.$^{ \rm { 6)}}$ Carter has demonstrated that the Hamilton-Jacobi equation for the
   geodesics in the Taub-NUT metric separates in specific coordinate systems.$^{ \rm { 7)}}$ The gravitomagnetic monopole source effects
   have been also examined in the Taub-NUT spacetime.$^{ \rm { 8)}}$ In the Kerr-Taub-NUT-de Sitter metrics, separability of
the Hamiltonian-Jacobi equation has been studied in higher dimensions.$^{ \rm { 9)}}$ Recently, a rotating Schwarzschild black hole has
been studied to investigate effective potentials for null and timelike geodesics of particles and hydrodynamics associated with general
 relativistic Euler equation for the steady state axisymmetric fluid.$^{ \rm { 10)}}$ Recently, calculations of energy, spatial momentum and angular momentum for some specific Kerr-NUT spacetimes have been done.$^{ \rm { 11)}}$

Hawking's ground-breaking work$^{ \rm { 12)}}$ on black hole evaporation and information loss
is based on the idea that a pair of particles is created just inside the event horizon
and from this pair the positive energy particle tunnels out of the hole and appears
as Hawking radiation. The negative energy particle tunnels inwards and results in
decrease of the mass of the black hole. The energy of the particle changes sign
as it crosses the horizon. These particles follow trajectories which cannot be explained
classically. This process of particle evaporation from black holes is thus a
phenomenon of quantum tunneling of particles through the event horizon.$^{ \rm { 13)}}$ This
contributes to the change in mass, angular momentum and charge of the black hole,
which change its thermodynamics as well. The particle travels in time so that the
action becomes complex and the dynamics of the outgoing particle is governed by the
imaginary part of this action. This action has been calculated using the radial null
geodesics$^{ \rm { 13)}}$ or the so-called Hamilton-Jacobi method$^{ \rm { 14)}}$ for various spacetimes. Using
the method of the radial null geodesics Hawking radiation as a tunneling process
has been analyzed for the Kerr and Kerr-Newman black holes.$^{ \rm { 15)}}$ They have done
the analysis using transformed forms of these metrics and they do not consider the
question of quantum corrections at all. The Hamilton-Jacobi method has been used
to calculate quantum corrections to the Hawking temperature and the Bekenstein-
Hawking area law for the Schwarzschild, anti-de Sitter Schwarzschild and Kerr black
holes.$^{ \rm { 14)}}$

At present, teleparallel theory seems to be popular
again, and there is a trend of analyzing the basic solutions
of general relativity (GR) with teleparallel theory and comparing
the results. The TEGR is a viable alternative geometrical description
of Einstein's GR written in terms of
the tetrad field.  It is considered as an essential part of generalized
non-Riemannian theories such as the Poincar$\acute{e}$  gauge
theory$^{ \rm { 16)}}$ or metric-affine gravity.$^{ \rm { 17)}}$ The physics
relevant to geometry may be related to the teleparallel
description of gravity.$^{ \rm { 16)}}$ Within the context of
metric-affine gravity, a stationary axially symmetric exact
solution of the vacuum field equations is obtained for
a specific gravitational Lagrangian by using prolongation
techniques (see \cite{BH} and references therein). A relation between spinor Lagrangian and
teleparallel theory is established.$^{ \rm { 19)}}$  In the framework of the TEGR
it has been possible to address the long-standing problem
of defining energy, momentum and angular momentum
of the gravitational field.$^{ \rm { 20)}}$ The tetrad field seems to
be a suitable field quantity to address this problem, because
it yields the gravitational field and at the same time
establishes a class of reference frames in spacetime.$^{ \rm { 21)}}$
Moreover, there are simple and clear indications that the
gravitational energy-momentum defined in the context of
the TEGR provides a unified picture of the concept of
mass-energy in special and general relativity. {\it It is the object of the current research to extend
 the previous work$^{ \rm { 11)}}$ by including other Kerr-NUT spacetimes to show if the previous procedure works correctly  or not?
  Also the physical quantities related to these tetrads, i.e, entropy, Hawking temperature are calculated.}\footnote{Many basic equations used in this calculation are reviewed explicitly in$^{ \rm { 11)}}$, and references therein. Therefore, in this work repetitions will be omitted.}

  In \S 2,  the first Kerr-NUT spacetime is given and calculation of its energy, spatial and angular momentum have been achieved
   using the definition
  of the gravitational energy-momentum, which is coordinate
independent, and  common know results are obtained.   In \S 3, the second Kerr-NUT spacetime is provided, that is linked
 to the first tetrad  through a local Lorentz transformation, and the conserved quantities, energy, spatial and angular momentums
  are calculated and a finite result is obtained.  In \S 4, it is shown by explicate calculation
 that the energy of the third Kerr-NUT spacetime using the definition
  of the gravitational energy-momentum is always divergent! To make the calculation clearer we use the Riemannian connection  1-form,
  ${\widetilde{\Gamma}_\alpha}^\beta$ and  repetition of
the calculation of energy has been done and  same divergent value is obtained. Therefore,
  a new local Lorentz
transformation is employed in \S5. When this transformation is applied to the Kerr-NUT spacetime, third one, and repetition of  the
calculation of energy either using the gravitational
energy-momentum or  the Riemannian connection  1-form, ${\widetilde{\Gamma}_\alpha}^\beta$ one can get a
finite and acceptable result. In \S 6, another Kerr-NUT spacetime is given, fourth one, calculation of
energy using the two definitions has been  achieved and  a divergent result is presented. In \S 7, using another local Lorentz transformation
 an acceptable and finite result is obtained.  In \S 8, we use the Euclidean continuation method to calculate the energy,
 entropy, temperature. Also we show the consistent of the first law of thermodynamics for Kerr-NUT spacetime and also show
 that the quantum statistical relation holds.  \S 9 is
devoted to main result and discussion.  In \S 10, many local Lorentz transformations, that
can be viewed as regularizing tools for the calculations of energy and momentum,  are provided.  \vspace{0.4cm}\\

\newsection{First Kerr-NUT spacetime}
The covariant form of the first Kerr-NUT tetrad field having  axial symmetry in
spherical coordinates $(t,r,\theta, \phi)$, can be written as
 \be
\left( {h^\alpha}_i \right)_1= \left( \matrix{
{\cal F}_1 & 0 & 0 & {\cal F}_2\vspace{3mm} \cr 0 & {\cal F}_3 & 0 & 0 \vspace{3mm}
\cr 0 & 0 & {\cal F}_4 & 0 \vspace{3mm}  \cr 0
 & 0 &0& -{\cal F}_5\sin\theta \cr } \right), \ee
 where  ${\cal F}_i$, $i=1\cdots 5$ are  functions of
 $r$ and $\theta$  having the form  \ba
{\cal F}_1 \A=\A \sqrt{\frac{{\cal D}}{\cal C}}, \qquad {\cal F}_2={\frac{\cal F}{\sqrt{\cal C D}}}, \qquad
  {\cal F}_3=\sqrt{\frac{{\cal C}}{{\cal D}_1}},  \qquad {\cal F}_4=\sqrt{\cal {C}}},   \qquad {\cal F}_5=\sqrt{\frac{{\cal C}{\cal D}_1}{{\cal D}},
 \nonu
\qquad {\cal C}\A=\A{r^2+\left(L+a\cos\theta\right)^2}, \qquad {\cal D } =
{r^2-2Mr+a^2\cos^2\theta-L^2}, \qquad {\cal D }_1 =
{r^2-2Mr+a^2-L^2}, \nonu
 {\cal F}\A=\A -2\Biggl(a[L^2+Mr]\sin^2\theta+{\cal D }_1L\cos\theta\Biggr),
   \ea  where  $M$ , $a$  and $L$ are the mass,  the rotation  and the NUT parameters respectively.$^{ \rm { 22)}}$ We consider a non asymptotically flat spacetime in this paper, and impose the boundary condition that for $r
\rightarrow \infty$ and $L\rightarrow 0$
 the tetrad (1) approaches the tetrad of Minkowski spacetime,  in Cartesian coordinate, i.e.,
  $O_{\alpha \beta}=diag(+1,-1,-1,-1)$.
 The metric tensor $g_{i j}\stackrel {\rm def.}{=}  O_{\mu \nu} {h^\mu}_i {h^\nu}_j$ associated with the tetrad field (1) has the form
 \be
ds^2= {{\cal F}_1}^2 dt^2
-{{\cal F}_3}^2  dr^2 -{{\cal F}_4}^2
d\theta^2-\left({{\cal F}_5}^2\sin^2\theta-{{\cal F}_2}^2\right) d\phi^2- {\cal F}_1{\cal F}_2 dt d\phi, \ee
which is the Kerr-NUT  spacetime written in
Boyer-Lindquist coordinates.$^{ \rm { 11)}}$

Now we are going to calculate the energy content of the tetrad
field  (1) using  (2). The non-vanishing components of the tensor $\Sigma^{a b c}$\footnote{${\Sigma}^{abc} \stackrel {\rm def.}{=} \displaystyle{1
\over 4}\left(T^{abc}+T^{bac}-T^{cab}\right)+\displaystyle{1 \over
2}\left(O^{ac}T^b-O^{ab}T^c\right)$ where  $T^{abc}$ and $T^a$
are the torsion tensor and the basic vector field.$^{ \rm { 11)}}$}  needed to the
calculation of energy have the form
\ba
 \Sigma^{(0) 0 1} \A \cong \A \frac{-\sin\theta}{2r^2 }\Biggl(2r^3-2Mr+2a^2M\cos^2\theta+ra^2\sin^2\theta+2(2M-r)aL\cos\theta-2rL^2\Biggr)+O\left(\frac{1}{r^3}\right),\nonu
 \Sigma^{(3) 0 1}\A \cong \A-\frac{r^2L\cos\theta+Mra\sin^2\theta-2aL\cos^2\theta(L+a\cos\theta)+L\cos\theta \ {\cal D}_1}{2r^2}+O\left(\frac{1}{r^3}\right).
\ea
Using Eq. (4), the energy associated with spacetime (1) takes the form
\ba
\A \A P^{(0)}=E=-\oint_{S \rightarrow \infty}
 dS_k \Pi^{(0) k}=-\displaystyle {1  \over 4 \pi} \oint_{S \rightarrow \infty}
 dS_k  \sqrt{-g}
{h^{(0)}}_\mu \left({\Sigma}^{\mu 0 k}-{{\Sigma}^{\mu 0 k}}_{M=0,a=0,L=0} \right)\nonu
\A \A \cong
M+\frac{L^2}{r}+O\left(\frac{1}{r^2}\right).\ea
It follows from Eq. (5) that
energy of the Kerr-NUT spacetime is finite and physically acceptable.
Using Eq. (4) one can get
 the spatial momentum in the
form \be P_1=-\oint_{S \rightarrow \infty}
 dS_k \Pi^{(1) k}=-\displaystyle {1  \over 4 \pi} \oint_{S \rightarrow \infty}
 dS_k  e {\Sigma}^{(1) 0 k} =0,\quad by \ same \  method
\ P_2=0, \quad P_3\cong\left(\frac{1}{r}\right).\ee

Now turn our attention to  calculate  the angular-momentum.
The non vanishing components of the
angular-momentum are given by \ba M^{(0)(1)}(r,\theta,\phi) \A \A \cong
 \frac{\sin\theta}{4\pi r}\{L^2+rM+aL\cos\theta-2r^2\}+O\left(\frac{1}{r^2}\right), \nonu
  M^{(0)(2)}(r,\theta,\phi) \A \A  \cong
 \frac{-1}{16\pi r}(2r^2\cos\theta-2La(2-3\cos^2\theta)-a^2\cos\theta[2-3\cos\theta])+O\left(\frac{1}{r^2}\right), \nonu
M^{(1)(3)}(r,\theta,\phi) \A \A   \cong \frac{-Ma\sin^2\theta}{4\pi r}+O\left(\frac{1}{r^2}\right),\quad
 M^{(2)(3)}(r,\theta,\phi) \cong \frac{[ML-rL+2Ma\cos\theta]\sin\theta}{4\pi r}+O\left(\frac{1}{r^2}\right).\nonu
 \A \A \ea
Using Eq. (7) we get
 \be L^{(0)(1)} = {\int_0^\pi}{\int_0^{2\pi}}{\int_{0}^\infty}
 d\theta d\phi dr \left[M^{(0)(1)} \right] =
0,\ee which is a consistence results.
  By  the same  method one can  obtain \be
 L^{(0)(2)} =L^{(1)(3)} = L^{(0)(3)} =
 L^{(1)(2)}= L^{(2)(3)}= 0.\ee Eq. (9) shows that the angular momentum associated with tetrad (1) not right and this may
 be  linked to the definition used in the calculation.$^{ \rm { 11)}}$
\newsection{Second Kerr-NUT spacetime}
The covariant form of the second Kerr-NUT tetrad field having  axial symmetry in
spherical coordinates, can be written as
 \ba
\left( {h^\alpha}_i \right)_2\A=\A \left( \matrix{ {\cal F}_1
& 0 & 0 &  {\cal F}_2 \vspace{3mm} \cr 0 & \sin\theta\cos\phi \  {\cal F}_3&  \cos\theta\cos\phi \ {\cal F}_4 & -\sin\theta\sin\phi \ {\cal F}_5 \vspace{3mm} \cr 0 & \sin\theta\sin\phi \ {\cal F}_3&  \cos\theta\sin\phi \ {\cal F}_4 &\sin\theta\cos\phi \ {\cal F}_5  \vspace{3mm} \cr  0 & \cos\theta \ {\cal F}_3&  -\sin\theta \ {\cal F}_4 & 0 \cr } \right)\nonu
\A \A \ \ where  \ \ {\cal F}_i, \  \ i=1 \cdots 5 \ \ are \ \ defined  \ \ by \ \  Eq. \ \ (13).  \ea
  Tetrad  (10) has the same associated metric of  tetrad (1), i.e., Kerr-NUT spacetime given by Eq. (3). Tetrad (10) is related to tetrad (1)
  through the relation $\left( {h^\alpha}_i \right)_2=\left({{\Lambda}^\alpha}_\gamma\right) \left( {h^\gamma}_i \right)_1$,
  where $\left({{\Lambda}^\alpha}_\gamma\right)$ is the local Lorentz transformation given by
   \be
\left({{\Lambda}^\alpha}_\gamma\right)  \stackrel {\rm def.}{=}
\left( \matrix{ 1 &  0 & 0 & 0 \vspace{3mm} \cr  0  &  \sin\theta
\cos\phi &  \cos\theta \cos\phi & - \sin\phi \vspace{3mm} \cr 0  &
\sin \theta \sin \phi& \cos\theta \sin\phi & \cos\phi \vspace{3mm}
\cr 0  & \cos\theta & -\sin\theta  & 0 \cr }\right)\;
.\ee
Following the same technics used in \S 2 to calculate the energy associated with tetrad (10), we finally get the non-vanishing components of $\Sigma^{a 0 1}$ up to order $O\left(\frac{1}{r^3}\right)$ as \ba
 \Sigma^{(0) 0 1} \A \cong \A \frac{-\sin\theta}{2r^2 }\Biggl(2r^3-2Mr+2a^2M(3\cos^2\theta-1)+ra^2\sin^2\theta+2(2M-r)aL\cos\theta-2rL^2\Biggr)+O\left(\frac{1}{r^3}\right),\nonu
 \Sigma^{(1) 0 1}\A \cong \A\frac{Ma\sin^2\theta\cos\phi }{2r}+O\left(\frac{1}{r^2}\right), \qquad \Sigma^{(2) 0 1} \cong \frac{Ma\sin^2\theta\sin\phi }{2r}+O\left(\frac{1}{r^2}\right).
\ea Using Eq. (12) to calculate  energy one can get the same value of  first tetrad as given by Eq. (5). Also the form of  spatial momentum components are the same as that of  Eq. (6). Finally, the angular momentum linked to tetrad (10) has the from (9).
\newsection{Third Kerr-NUT spacetime}
The covariant form of the third Kerr-NUT tetrad  can be written as
 \ba
\left( {h^\alpha}_i \right)_3\A=\A \left( \matrix{ -\sqrt{\displaystyle\frac{{\cal D}_1}{\cal C}}
& 0 & 0 &  \sqrt{\displaystyle\frac{{\cal D}_1}{\cal C}}\left(a\sin^2\theta-2L\cos\theta\right) \vspace{3mm} \cr 0 & \displaystyle\frac{-r\sin\theta}{  \sqrt{\cal D}_1}&   {\cal C}_1 & 0
 \vspace{3mm} \cr \displaystyle\frac{a\sin\theta}{\sqrt{\cal C}} & 0&  0 & \displaystyle\frac{-\sin\theta(r^2+a^2+L^2)}{\sqrt{\cal C}}
   \vspace{3mm} \cr  0 &  \displaystyle\frac{{\cal C}_1}{\sqrt{{\cal D}_1}} &  -r\sin\theta & 0 \cr } \right),  \ea
where ${\cal C}_1=\sqrt{{\cal C}-r^2\sin^2\theta}$.
  Tetrad (13) has the same associated metric of  tetrad (1).  Tetrad (13) is related to tetrad (1)
  through the relation $\left( {h^\alpha}_i \right)_3=\left({{\Lambda_1}^\alpha}_\gamma\right) \left( {h^\gamma}_i \right)_1$
  where $\left({{\Lambda_1}^\alpha}_\gamma\right)$ is another local Lorentz transformation given by
   \be
\left({{\Lambda_1}^\alpha}_\gamma\right)  \stackrel {\rm def.}{=}
\left( \matrix{ \sqrt{\displaystyle\frac{{\cal D}_1}{{\cal D}}}  &  0 & \displaystyle\frac{a\sin\theta}{\sqrt{{\cal D}}} & 0 \vspace{3mm} \cr  0  & \displaystyle\frac{r\sin\theta}{\sqrt{{\cal C}}}  &  0 & \displaystyle\frac{{\cal C}_1}{\sqrt{{\cal C}}}  \vspace{3mm} \cr 0  &
\displaystyle\frac{{\cal C}_1}{\sqrt{{\cal C}}}&0 & -\displaystyle\frac{r\sin\theta}{\sqrt{{\cal C}}}  \vspace{3mm}
\cr \displaystyle\frac{a\sin\theta}{\sqrt{{\cal D}}}   & 0 &  \sqrt{\displaystyle\frac{{\cal D}_1}{{\cal D}}}&0  \cr }\right)\;
,\ee
Following the same technics used in \S 2 and \S 3 to calculate energy one can get the non-vanishing components of $\Sigma^{a 0 1}$ asymptotically as\ba
 \Sigma^{(0) 0 1} \A \cong \A \frac{-\sin\theta}{4r\cos^2\theta }\Biggl(6r^2\cos^2\theta-7a^2\cos^2\theta\sin^2\theta+4Mr\cos^2\theta+L^2(1+3\cos^2\theta)+12aL\cos^3\theta\Biggr)+O\left(\frac{1}{r^2}\right),\nonu
 \Sigma^{(2) 0 1}\A \cong \A \frac{[Ma-3ar]\sin^2\theta+4L\cos\theta[r-M]}{4r}+O\left(\frac{1}{r^2}\right).
\ea
Using Eq. (15)  we finally obtain
\ba
\A \A P^{(0)}=E=-\oint_{S \rightarrow \infty}
 dS_k \Pi^{(0) k}=-\displaystyle {1  \over 4 \pi} \oint_{S \rightarrow \infty}
 dS_k  \sqrt{-g}
{h^{(0)}}_\mu {\Sigma}^{\mu 0 k}  \cong \infty!\ea
It follows from Eq. (16) that the
energy of the Kerr-NUT spacetime is divergent and this is not an acceptable.

Due to the above non physical result and to make the picture  clearer  another method will be use
 to calculate the energy of tetrad (13) to show if the divergent result will continue or not?

The coframe of tetrad  (13), is described by the components\footnote{$\vartheta^{\alpha}=h^\alpha_i
dx^i$. We use the relativistic units in this calculation. $
\widetilde{H}_\alpha=\frac{1}{2\kappa}{\widetilde{\Gamma}}^{\beta
\gamma}\wedge  \eta_{\alpha \beta \gamma}, \qquad
{\Gamma_\alpha}^\beta \stackrel {\rm def.}{=} {\widetilde
{\Gamma}_\alpha}^\beta -{K_\alpha}^\beta$ with ${\widetilde
{\Gamma}_\alpha}^\beta $  is the purely Riemannian connection and
$K^{\mu \nu}$ is the contorsion 1-form}:
 \ba
{\vartheta_1}^{\hat{0}}\A=\A -\sqrt{\frac{{\cal D}_1}{\cal C}}\left([a\cos^2\theta+2L\cos\theta]d\phi+dt\right), \qquad
{\vartheta_1}^{\hat{1}}=-\frac{r\sin\theta dr-\sqrt{\cal D}_1{\cal C}_1 d\theta}{\sqrt{{\cal D}_1}},\nonu
 {\vartheta_1}^{\hat{2}} \A=\A -\frac{\sin\theta}{\sqrt{{\cal C}}}\left([r^2+L^2+a^2]d\phi-adt\right)
, \qquad  {\vartheta_1}^{\hat{3}}=-\frac{{\cal C}_1dr+r\sin\theta \sqrt{\cal D}_1 d\theta}{\sqrt{{\cal D}_1}}. \ea
   Taking coframe  (17), as
well as the Riemannian connection
${\tilde{\Gamma}_\alpha}^\beta$  and substitute into the translational momentum,  $\widetilde{H}_\alpha$, we finally get\footnote{$\cdots$ means terms which are multiply by $d\theta\wedge dr$, $d\theta\wedge dt$, $dr\wedge d\phi$ etc.}
\ba \widetilde{H}_{\hat{0}} \A \cong \A \frac{- \sin\theta }{8r\pi \cos^2\theta} \Biggl[2r^2\cos^2\theta+5a^2\cos^2\theta
-L^2\sin^2\theta-8aL\cos^3\theta-5a^2\cos^4\theta\Biggr]d\theta\wedge d\phi
 +\cdots+O\left(\frac{1}{r^2}\right).\nonu
 \A \A \ea
Using Eq. (18) to compute the total energy at a fixed time in the 3-space with
a spatial  2-dimensional boundary surface $\partial S =\{r = R,
\theta,\phi\}$ we finally obtain  \be \widetilde{E} =\int_{\partial S}
\widetilde{H}_{\hat{0}}=\infty!\ee which is identical with Eq. (16). Eqs. (16) and (19) show that the two definitions,
the gravitational energy-momentum and the translational momentum give equal form of energy which is not acceptable. This means that tetrad (13)
is inconvenient one and should multiply by some appropriate local Lorentz transformation to bypass the above inconsistence result.

\newsection{On the choice of the frame}
Let us consider the local Lorentz transformation described by the matrix
 \be \left({{\Lambda_2}^\alpha}_\beta\right)
= \left( \matrix{ H_1 &  H_2 &
H_3 &  H_4
 \vspace{3mm} \cr  K_1 \sin\theta
\cos\phi  & K_2 \sin\theta \cos\phi & K_3 \cos\theta \cos\phi
 & K_4 \sin\phi  \sin\theta \vspace{3mm} \cr
  L_1 \sin\theta \sin\phi  & L_2  \sin\theta \sin\phi &L_3
   \cos\theta
\sin\phi & L_4 \cos\phi \sin\theta \vspace{3mm} \cr
 N_1\cos\theta&   N_2 \cos\theta & N_3\sin\theta  &
   N_4 \cos\theta  \cr }
\right)\; , \ee where $H_i\; ,  \ K_i\; , \ L_i \ \ and \ \  N_i\; ,
\
 i=1 \cdots 4 $ are defined as:
 \ba H_1 \A
=\A -\frac{r^2-Mr+a^2+aL\cos\theta}{\sqrt{{\cal C }{\cal D}_1}} , \qquad
H_2=-\frac{(Mr+L^2+aL\cos\theta)r\sin\theta}{{\cal C }\sqrt{{\cal D}_1}}\; ,
 \quad H_3=-\frac{a\sin\theta}{\sqrt{{\cal C}}}\; , \nonu
 H_4 \A =\A -\frac{(Mr+L^2+aL\cos\theta){\cal C}_1}{{\cal C }\sqrt{{\cal D}_1}}\; ,\qquad
  K_1 = -\displaystyle{(Mr+L^2+aL\cos\theta-a^2)\cos\phi+ar_1\sin\phi \over {\cos\phi \ \sqrt{{\cal C }{\cal D}_1}}}\; ,\nonu
  K_2 \A =\A -\frac{r[(r^2-Mr+aL\cos\theta)\cos\phi+ar_1\sin\phi]\sin^2\theta-\alpha\cos\theta{\cal C}_1\sqrt{{\cal D }_1}}
  {\cos\phi\sin\theta \sqrt{{\cal D}_1}{\cal C } } \; ,\nonu
  K_3 \A =\A -\displaystyle{\beta \over \cos\theta\cos \phi \sqrt{\cal C}} \; , \qquad
   K_4 =  -\frac{{\cal C}_1[(r^2+aL\cos\theta-Mr)\cos\phi+ar_1\sin\phi]+r\alpha \cos\theta\sqrt{{\cal D}_1}}
  {\sin\phi \sqrt{{\cal D}_1}{\cal C }}\; , \nonu
 L_1 \A=\A -\displaystyle{(Mr+L^2+aL\cos\theta-a^2)\sin\phi-ar_1\cos\phi \over {\sqrt{{\cal C }{\cal D}_1}}\sin\phi}\; ,\nonu
 L_2 \A =\A -\frac{r[(r^2-Mr+aL\cos\theta)\sin\phi-ar_1\cos\phi]\sin^2\theta-\beta\cos\theta {\cal C}_1\sqrt{{\cal D }_1}}
  {\sin\phi\sin\theta\sqrt{{\cal D}_1}{\cal C }} \; ,\quad
    L_3 = \displaystyle{\alpha \over \cos\theta \sin \phi \sqrt{\cal C}} \; , \nonu
  L_4 \A=\A  -\frac{{\cal C}_1[(r^2+aL\cos\theta-Mr)\sin\phi-ar_1\cos\phi]+r\beta \cos\theta\sqrt{{\cal D}_1}}
  {\cos\phi \sqrt{{\cal D}_1}{\cal C }}\; , \qquad
   N_1 =-\displaystyle{(Mr+L^2+aL\cos\theta) \over {\sqrt{{\cal C }{\cal D}_1}}}\; ,\nonu
 N_2 \A =\A -\frac{\sin\theta[r\cos\theta(r^2+a^2-Mr+aL\cos\theta)+r_1{\cal C}_1\sqrt{{\cal D}_1}]}{{\cos\theta \ {\cal C}\sqrt{{\cal D}_1}}} \; ,  \qquad \qquad N_3=0 \nonu
  N_4 \A=\A - \frac{\cos\theta(r^2+a^2-Mr+aL\cos\theta){\cal C}_1-rr_1\sqrt{{\cal D}_1}\sin^2\theta}{{\cos\theta \ {\cal C}\sqrt{{\cal D}_1}}}\; , \nonu
 \A \A \ea
 where $\ \alpha, \ \  \beta, \ \ and \ \ r_1$ are defined by
  \ba \alpha  \stackrel
{\rm def.}{=} r_1\cos \phi+a \sin\phi, \qquad  \qquad  \beta \stackrel
{\rm def.}{=} r_1
\sin \phi-a \cos \phi\;, \qquad  \qquad r_1 \stackrel
{\rm def.}{=} \sqrt{r^2+L(L+2a\cos\theta)}.\nonu
 \A \A \ea
From Eqs. (13) and (20) we construct the new tetrad \footnote{  Eq. (23) is an exact solution to the equation of motion of TEGR. This case is studied intensively by  Hayashi and Shirafuji (cf., Ref. 13) Eqs. (7$\cdot$ 2)$\sim$(7$\cdot$ 11) and references therein).}
\be
\left( {h^\alpha}_i \right)_4
=\left({{\Lambda_2}^\alpha}_\gamma \right)
 \left( {h^\gamma}_i \right)_3. \ee
  Using Eq. (23) to calculate the non-vanishing components needed to the calculations of energy,  we finally get\footnote{$\Sigma^{(a) b c}=\left( {h^a}_\alpha \right)_4\Sigma^{\alpha b c}.$ \ \  Terms like $aML$, $Ma^2$, $Ma^2L$, etc. $\cdots$ are neglected in these calculations.}
 \ba {\Sigma}^{(0) 0 1} \A \cong \A \frac{\sin\theta(Mr^2+rL^2-raL\cos\theta)}{r^2}, \nonu
 {\Sigma}^{(1) 0 1}\A=\A {\Sigma}^{(2) 0 1}={\Sigma}^{(3) 0 1} \cong \frac{-\sin\theta(r^3-2Mr^2-2rL^2-2raL\cos\theta)}{r^2}. \ea
 From Eq. (24) one can obtain the energy of Eq. (23) in the from \footnote{We introduce ${{\Sigma}^{\mu 0 k}}_{M=0,a=0,L=0}$ to remove the
divergence appearers from term like $r$. It is worth to mention that we cannot  use the expression ${{\Sigma}^{\mu 0 k}}_{r \rightarrow \infty}$ because the spacetime we use is not asymptotically flat.}
\ba
\A \A P^{(0)}=E=-\oint_{S \rightarrow \infty}
 dS_k \Pi^{(0) k}=-\displaystyle {1  \over 4 \pi} \oint_{S \rightarrow \infty}
 dS_k  \sqrt{-g}
{h^{(0)}}_\mu \left({\Sigma}^{\mu 0 k}-{{\Sigma}^{\mu 0 k}}_{M=0,a=0,L=0} \right)\nonu
\A \A \cong M+\frac{L^2}{r}-\frac{L^2M}{r^2}+O\left(\frac{1}{r^3}\right).\ea
Eq. (25) is a satisfactory results.$^{ \rm { 22)}}$  Using Eq. (24) one can get the spatial momentum related to tetrad (23) to has
 the same form of Eq. (6).\footnote{Terms like $M^2$, $L^3$, $L^3M$,$ M^2a$, $\cdots$ etc. are neglected in this calculations.}

Now turn our attention to  calculate  the angular-momentum related to tetrad (23).
The non vanishing components needed  to compute the
angular-momentum up to order $O\left(\frac{1}{r^2}\right)$ are given by \ba M^{(0)(1)}(r,\theta,\phi) \A \A \cong
 \frac{-1}{8\pi r}(\{2r^2+2aL\cos^3\theta+L^2\cos^2\theta-2[L^2+rM+a^2]\sin^2\theta-4aL\sin^2\theta\cos\theta\}\cos\phi \nonu
 \A \A +ar\sin^2\theta\sin\phi)+O\left(\frac{1}{r^2}\right), \nonu
  M^{(0)(2)}(r,\theta,\phi) \A \A  \cong
\frac{-1}{8\pi r}\biggl(\{2r^2+2aL\cos^3\theta+L^2\cos^2\theta-2[L^2+rM+a^2]\sin^2\theta-4aL\sin^2\theta\cos\theta\}\sin\phi, \nonu
 \A \A +ar\sin^2\theta\cos\phi \Biggr)+O\left(\frac{1}{r^2}\right),\nonu
M^{(0)(3)}(r,\theta,\phi) \A \cong \A  \frac{\sin\theta(6aL\cos^2\theta+2Mr\cos\theta+3L^2\cos\theta-2aL)}{8\pi r}+O\left(\frac{1}{r^2}\right),\nonu
M^{(1)(2)}(r,\theta,\phi)\A \A \cong \frac{\sin\theta(3aM\cos^2\theta+2ML\cos\theta-2rL\cos\theta-aM)}{8\pi r}+O\left(\frac{1}{r^2}\right),\nonu
M^{(1)(3)}(r,\theta,\phi) \A \A   \cong \frac{-\sin^2\theta([3aM\cos\theta-2rL+2LM]\sin\phi+aL\cos\phi)}{8\pi r}+O\left(\frac{1}{r^2}\right),\nonu
 M^{(2)(3)}(r,\theta,\phi) \A \A \cong\frac{\sin^2\theta([3aM\cos\theta-2rL+2LM]\cos\phi-aL\sin\phi)}{8\pi r}+O\left(\frac{1}{r^2}\right).\ea
Using Eq. (26) we obtain the components  of the angular momentum linked to tetrad (23) same as that  provided by Eqs. (8) and (9).

We show by explicit calculations that the energy-momentum tensor,
that is a coordinate independent,   gives   inconsistent
result of the angular momentum when applied to the tetrad field
given by Eq. (23)!

Using the   translational momentum employed in the previous section to show if tetrad (23) to continue providing  the consistence physical results of energy and spatial momentum or not.  The coframe related to tetrad (23) is
described by the components:
\ba {\vartheta_4}^{\hat{0}}\A=\A \frac{1}{{\cal D}_1{\cal C}} \Biggl\{{\cal D}_1[r^2-Mr+a\cos\theta(L+a\cos\theta)]dt+{\cal C}[Mr+L(L+a\cos\theta)]dr\nonu
\A \A +{\cal D}_1[a^2L\cos^3\theta-arM\cos^2\theta+aL^2\cos^2\theta+2r^2L\cos\theta-2rML\cos\theta
+a^2L\cos\theta+arM+aL^2]d\phi \Biggr\},\nonu
 {\vartheta_4}^{\hat{1}}\A=\A \frac{1}{{\cal D}_1{\cal C}} \Biggl\{{\cal D}_1[Mr+L(L+a\cos\theta)]\sin\theta \cos\phi dt+{\cal C}\{[r^2-Mr+aL\cos\theta]\cos\phi+ar_1\sin\phi\}\sin\theta dr\nonu
 \A \A +{\cal C}{\cal D}_1\alpha \cos\theta d\theta+
{\cal D}_1[r_1\sin\phi{\cal C}-\cos\phi\{ar^2+2aL^2-3aL^2\cos^2\theta-2L^3\cos\theta-a^2L\cos^3\theta\nonu
\A \A +3a^2L\cos\theta+a^3\cos^2\theta
-arM\cos^2\theta-2rML\cos\theta+raM\}] \sin\theta d\phi\Biggr\},\nonu
 {\vartheta_4}^{\hat{2}}\A=\A \frac{1}{{\cal D}_1{\cal C}} \Biggl\{{\cal D}_1[Mr+L(L+a\cos\theta)]\sin\theta \sin\phi dt+{\cal C}\{[r^2-Mr+aL\cos\theta]\sin\phi-ar_1\cos\phi\}\sin\theta dr\nonu
 \A \A +{\cal C}{\cal D}_1\beta \cos\theta d\theta-
{\cal D}_1[r_1\cos\phi{\cal C}+\sin\phi\{ar^2+2aL^2-3aL^2\cos^2\theta-2L^3\cos\theta-a^2L\cos^3\theta\nonu
\A \A +3a^2L\cos\theta+a^3\cos^2\theta
-arM\cos^2\theta-2rML\cos\theta+raM\}] \sin\theta d\phi\Biggr\},\nonu
 {\vartheta_4}^{\hat{3}}\A=\A \frac{1}{{\cal D}_1{\cal C}} \Biggl\{{\cal D}_1[Mr+L(L+a\cos\theta)]\cos\theta dt+{\cal C}\{[r^2-Mr+aL\cos\theta]\}\cos\theta dr\nonu
 \A \A -{\cal C}{\cal D}_1r_1 \sin\theta d\theta+
{\cal D}_1[2L\cos\theta-a\sin^2\theta][Mr+L(L+a\cos\theta)] \cos\theta d\phi\Biggr\}.\ea

  If we take coframe  (27), as
well as the trivial Weitzenb$\ddot{o}$ck connection, i.e.,
${\Gamma^\alpha}_\beta=0$   we finally reach  the
temporal component of the translation momentum in the form
\ba \widetilde{H}_{\hat{0}} \A \A \cong -\frac{\sin \theta}{8r^2\pi} \left[3a^2M\cos^2\theta+6aLM \cos\theta-2arL\cos\theta+2ML^2-2rL^2-a^2M+2r^3-2r^2M
\right]d\theta\wedge d\phi\nonu
\A \A +\cdots+O\left(\frac{1}{r^3}\right).\ea Computing the total energy up to order $O\left(\frac{1}{r^3}\right)$ at a
fixed time in the 3-space with a spatial  2-dimensional
boundary surface $\partial S =\{r = R, \theta,\phi\}$ we obtain\footnote{We introduce $\Biggl\{\widetilde{H}_{\hat{0}}\Biggr\}_{M=0,a=0,L=0}$ to remove the
divergence appearers from term like $r$. It is worth to mention that we cannot  use the expression $\Biggl\{\widetilde{H}_{\hat{0}}\Biggr\}_{r \rightarrow \infty}$ because the spacetime we use is not asymptotically flat.} \be
\widetilde{E} =\int_{\partial S} \left(\widetilde{H}_{\hat{0}}-\Biggl\{\widetilde{H}_{\hat{0}}\Biggr\}_{M=0,a=0,L=0}\right) \cong M+\frac{L^2}{R}-\frac{L^2M}{R^2}
+O\left(\frac{1}{R^3}\right),\ee
which is the energy of Kerr black hole when $L=0$.$^{ \rm { 22)}}$

 The  necessary components needed to calculate the spatial momentum
 $\widetilde{H}_{\hat{\alpha}}, \
\hat{\alpha}=1,2,3$ have the following  components  \ba \widetilde{H}_{\hat{1}}
\A=\A \frac{2\cos\phi \sin^2\theta[3aL\cos\theta+2L^2+2Mr+4M^2]}{r}d\theta\wedge
d\phi+\cdots, \nonu
\widetilde{H}_{\hat{2}} \A=\A \frac{2\sin\phi \sin^2\theta[3aL\cos\theta+2L^2+2Mr+4M^2]}{r}d\theta\wedge
d\phi+\cdots\nonu
\widetilde{H}_{\hat{3}}\A=\A  \frac{2\sin
\theta\left([3aL\cos\theta+2\{L^2+Mr\}\cos\theta-aL+4M^2]\cos\theta-aL\right)}{r}d\theta\wedge
d\phi+\cdots. \ea Using Eq. (30)  we finally get the spatial momentum in
the form \be P_1=P_2=P_3\cong O\left(\frac{1}{R^2}\right).\ee
\newsection{Fourth Kerr-NUT spacetime}
The covariant form of the fourth Kerr-NUT tetrad  can be written as
 \ba
\left( {h^\alpha}_i \right)_4\A=\A \left( \matrix{ -\sqrt{\displaystyle\frac{{\cal D}_1}{\cal C}}
& 0 & 0 &  \sqrt{\displaystyle\frac{{\cal D}_1}{\cal C}}\left(a\sin^2\theta-2L\cos\theta\right) \vspace{3mm} \cr  \displaystyle\frac{-a\sin\theta \sin\phi}{  \sqrt{\cal C}} & \displaystyle\frac{-r\sin\theta \cos\phi}{  \sqrt{\cal D}_1}&   {\cal C}_1 \cos\phi &  \displaystyle\frac{(r^2+a^2+L^2)\sin\theta \sin\phi}{  \sqrt{\cal C}}
 \vspace{3mm} \cr \displaystyle\frac{a\sin\theta \cos\phi}{  \sqrt{\cal C}}&\displaystyle\frac{-r\sin\theta \sin\phi}{  \sqrt{\cal D}_1}&  {\cal C}_1 \sin\phi &-\displaystyle\frac{(r^2+a^2+L^2)\sin\theta \cos\phi}{  \sqrt{\cal C}}
   \vspace{3mm} \cr  0 &  -\displaystyle\frac{{\cal C}_1}{\sqrt{{\cal D}_1}} &  -r\sin\theta & 0 \cr } \right).\nonu
   \A \A   \ea

  Tetrad  (32) has the same associated metric of  tetrad (1). Tetrad (32) is related to tetrad (1)
  through the relation $\left( {h^\alpha}_i \right)_5=\left({{\Lambda_3}^\alpha}_\gamma\right) \left( {h^\gamma}_i \right)_1$
  where $\left({{\Lambda_3}^\alpha}_\gamma\right)$ is the local Lorentz transformation given by
   \be
\left({{\Lambda_3}^\alpha}_\gamma\right)  \stackrel {\rm def.}{=}
\left( \matrix{ \sqrt{\displaystyle\frac{{\cal D}_1}{{\cal D}}}  &  -\displaystyle\frac{a\sin\theta\sin\phi}{\sqrt{{\cal D}}}  & \displaystyle\frac{a\sin\theta\cos\phi}{\sqrt{{\cal D}}} & 0 \vspace{3mm} \cr -\displaystyle\frac{a\sin\theta\sin\phi}{\sqrt{{\cal D}}}  & -\displaystyle\frac{L_5}{\sqrt{{\cal C D}}}  &  -\displaystyle\frac{L_6\cos\phi\sin\phi}{\sqrt{{\cal C D}}}  & -\displaystyle\frac{L_7\cos\phi\sin\theta}{\sqrt{{\cal C}}}  \vspace{3mm} \cr \displaystyle\frac{a\sin\theta\cos\phi}{\sqrt{{\cal D}}}  & -\displaystyle\frac{L_6\sin\phi\cos\phi}{\sqrt{{\cal C D}}}  &  -\displaystyle\frac{L_8}{\sqrt{{\cal C D}}}  & -\displaystyle\frac{L_7\sin\phi\sin\theta}{\sqrt{{\cal C}}}  \vspace{3mm}
\cr 0   & -\displaystyle\frac{L_7\cos\phi\sin\theta}{\sqrt{{\cal C}}}&  -\displaystyle\frac{L_7\sin\phi\sin\theta}{\sqrt{{\cal C}}}&-\displaystyle\frac{{\cal C}_1\cos\theta+r\sin^2\theta}{\sqrt{{\cal C}}}  \cr }\right)\;
,\ee
where \ba
L_5\A=\A \sqrt{{\cal D}}[r\sin^2\theta-{\cal C}_1 \cos\theta]\cos^2\phi-\sqrt{{\cal C}{\cal D}_1}\sin^2\phi,\qquad L_6=-\sqrt{{\cal D}}[r\sin^2\theta-{\cal C}_1 \cos\theta]+\sqrt{{\cal C}{\cal D}_1}\; ,\nonu
\quad L_7\A=\A({\cal C}_1+r\cos\theta)\; ,\qquad L_8=\sqrt{{\cal D}}[r\sin^2\theta-{\cal C}_1 \cos\theta]\sin^2\phi-\sqrt{{\cal C}{\cal D}_1}\cos^2\phi\; .\ea

Following the same technics used in the previous sections to calculate energy, we finally get the non-vanishing components of $\Sigma^{a 0 1}$ asymptotically as\ba
 \Sigma^{(0) 0 1} \A \cong \A \frac{-\sin\theta}{4r\cos^2\theta }\Biggl(6r^2\cos^2\theta-7a^2\cos^2\theta\sin^2\theta+4Mr\cos^2\theta+L^2(1+3\cos^2\theta)+12aL\cos^3\theta\Biggr)+O\left(\frac{1}{r^2}\right),\nonu
 \Sigma^{(2) 0 1}\A \cong \A \frac{[Ma-3ar]\sin^2\theta+4L\cos\theta[r-M]}{4r}+O\left(\frac{1}{r^2}\right).
\ea
Using Eq. (35),  we finally obtain the energy of tetrad (32) in the form
\ba
\A \A P^{(0)}=E=-\oint_{S \rightarrow \infty}
 dS_k \Pi^{(0) k}=-\displaystyle {1  \over 4 \pi} \oint_{S \rightarrow \infty}
 dS_k  \sqrt{-g}
{h^{(0)}}_\mu {\Sigma}^{\mu 0 k}  \cong \infty!\ea

 Using the translational momentum to recalculate the energy of tetrad (32) whose  coframe is described by the
 components:
 \ba
{\vartheta_1}^{\hat{0}}\A=\A -\sqrt{\frac{{\cal D}_1}{\cal C}}\left([a\cos^2\theta+2L\cos\theta]d\phi+dt\right), \qquad
{\vartheta_1}^{\hat{1}}=-\frac{\left(r\sin\theta dr-\sqrt{\cal D}_1\sqrt{{\cal C}-r^2\cos^2\theta}d\theta\right)}{\sqrt{{\cal D}_1}},\nonu
 {\vartheta_1}^{\hat{2}} \A=\A -\frac{\sin\theta}{\sqrt{{\cal C}}}\left([r^2+L^2+a^2]d\phi-adt\right)
, \qquad  {\vartheta_1}^{\hat{3}}=-\frac{\left(\sqrt{{\cal C}-r^2\cos^2\theta}dr+r\sin\theta \sqrt{\cal D}_1 d\theta\right)}{\sqrt{{\cal D}_1}}. \ea
   Taking coframe  (37), as
well as the Riemannian connection
${\tilde{\Gamma}_\alpha}^\beta$  and substitute into the translational momentum,  $\widetilde{H}_\alpha$, we finally get
\ba \widetilde{H}_{\hat{0}} \A \cong \A \frac{- \sin\theta }{8r\pi \cos^2\theta} \Biggl[2r^2\cos^2\theta+5a^2\cos^2\theta
-L^2\sin^2\theta-8aL\cos^3\theta-5a^2\cos^4\theta\Biggr]d\theta\wedge d\phi
 +\cdots+O\left(\frac{1}{r^2}\right).\nonu
 \A \A \ea
Using Eq. (38) to compute the total energy at a fixed time in the 3-space with
a spatial  2-dimensional boundary surface $\partial S =\{r = R,
\theta,\phi\}$ we finally obtain  \be \widetilde{E} =\int_{\partial S}
\widetilde{H}_{\hat{0}}=\infty!\ee which is identical with Eq. (36).

\newsection{On the choice of the frame}
Let us consider the local Lorentz transformation described by the matrix
\be \left({{\Lambda_4}^\alpha}_\beta\right)
= \left( \matrix{ H_1 &  H_5 &
H_6 &  H_4
 \vspace{3mm} \cr  K_1 \sin\theta
\cos\phi  & K_5 \sin\theta \cos\phi & K_6 \cos\theta \cos\phi
 & K_4 \sin\phi  \sin\theta \vspace{3mm} \cr
  L_1 \sin\theta \sin\phi  & L_5  \sin\theta \sin\phi &L_6
   \cos\theta
\sin\phi & L_4 \cos\phi \sin\theta \vspace{3mm} \cr
 N_1\cos\theta&   N_5 \cos\theta & N_6\sin\theta  &
   N_7 \cos\theta  \cr }
\right)\; , \ee where $H_i\; ,  \ K_i\; , \ L_i \ \ and \ \  N_i\; ,
\
 i=5 \ \ , \ \ 6$ and $N_7$ are defined as:

\ba
H_5\A=\A-\frac{\sin\theta[(Mr+L^2+aL\cos\theta)r\cos\phi-a\sin\phi\sqrt{{\cal C }{\cal D}_1}]}{{\cal C }\sqrt{{\cal D}_1}}\; ,\nonu
 H_6\A=\A-\frac{\sin\theta[(Mr+L^2+aL\cos\theta)r\sin\phi+a\cos\phi\sqrt{{\cal C }{\cal D}_1}]}{{\cal C }\sqrt{{\cal D}_1}}\; , \nonu
 K_5 \A =\A \frac{-1} {\cos\phi\sin\theta \ {\cal C }\ \sqrt{{\cal D}_1} }\Biggl\{(r_1\sqrt{{\cal D}_1{\cal C }}+r[r^2+La\cos\theta-rM]\sin^2\theta-r_1  {\cal C}_1 \cos\theta \ \sqrt{{\cal D }_1})\cos^2\phi\nonu
  \A \A +[rr_1a\sin^2\theta+a\sqrt{{\cal C}{\cal D }_1}-a \ {\cal C}_1 \ \cos\theta\sqrt{{\cal D}_1}]\cos\phi\sin\phi -r_1  \sqrt{{\cal C} {\cal D}_1}\Biggr\}
  \; ,\nonu
  K_6 \A =\A \frac{-1} {\cos\phi\cos\theta \ {\cal C } \ \sqrt{{\cal D}_1} }\Biggl\{(r_1\sqrt{{\cal D}_1{\cal C }}+r[r^2+La\cos\theta-rM]\sin^2\theta-r_1\cos\theta \ {\cal C}_1\ \sqrt{{\cal D }_1})\cos\phi\sin\phi \nonu
  \A \A -[rr_1a\sin^2\theta+a \sqrt{{\cal C}{\cal D }_1}-a \ {\cal C}_1 \ \cos\theta\sqrt{{\cal D}_1}]\cos^2\phi -a \ {\cal C}_1 \ \cos\theta\sqrt{{\cal D}_1}+arr_1\sin^2\theta\Biggr\}
  \; , \nonu
 L_5 \A =\A\frac{-1} {\cos\phi\sin\theta \ {\cal C } \ \sqrt{{\cal D}_1} }\Biggl\{(r_1\sqrt{{\cal D}_1{\cal C }}+r[r^2+La\cos\theta-rM]\sin^2\theta-r_1 {\cal C}_1 \cos\theta\sqrt{{\cal D }_1})\cos\phi \sin\phi\nonu
  \A \A -[rr_1a\sin^2\theta+a\sqrt{{\cal C}{\cal D }_1}-a \ {\cal C}_1 \cos\theta\sqrt{{\cal D}_1}]\cos^2\phi +a\sqrt{{\cal C}{\cal D}_1}\Biggr\}
  \; ,\nonu
  L_6 \A=\A \frac{1} {\cos\phi\cos\theta \ {\cal C } \ \sqrt{{\cal D}_1} }\Biggl\{(r_1\sqrt{{\cal D}_1{\cal C }}+r[r^2+La\cos\theta-rM]\sin^2\theta-r_1\ {\cal C}_1\ \cos\theta\sqrt{{\cal D }_1})\cos^2\phi \nonu
  \A \A +[rr_1a\sin^2\theta+a\sqrt{{\cal C}{\cal D }_1}-a \ {\cal C}_1\ \cos\theta\sqrt{{\cal D}_1}]\cos\phi\sin\phi+r_1 \ {\cal C}_1\ \cos\theta\sqrt{{\cal D}_1}-r[r^2+aL\cos\theta-rM]\sin^2\theta\Biggr\}
  \; , \nonu
   N_5 \A =\A -\frac{\sin\theta\cos\phi[r\cos\theta(r^2+a^2-Mr+aL\cos\theta)+r_1 \ {\cal C}_1\ \sqrt{{\cal D}_1}]}{{\cos\theta
   {\cal C}\sqrt{{\cal D}_1}}} \; , \nonu
   N_6 \A =\A -\frac{\sin\phi[r\cos\theta(r^2+a^2-Mr+aL\cos\theta)+r_1 {\cal C}_1 \sqrt{\cal D}_1]}{{ {\cal C} \sqrt{\cal D}_1}} \; . \nonu
   N_7 \A =\A -\frac{[\cos\theta(r^2+a^2-Mr+aL\cos\theta) {\cal C}_1 +rr_1\sin^2\theta \sqrt{\cal D}_1]}{{ {\cal C} \sqrt{\cal D}_1}} \; . \nonu
 \A \A \ea
From Eqs. (32) and (40) one can get
\be
\left( {h^\alpha}_i \right)_6
=\left({{\Lambda_4}^\alpha}_\gamma \right)
 \left( {h^\gamma}_i \right)_5. \ee
 Using  Eq. (42) to calculate the non-vanishing components needed to the calculations of energy,  we finally get
 \ba
 \Sigma^{(0) 0 1} \A \cong \A \frac{-\sin\theta}{r }\Biggl(r^2-Mr-L^2-La\cos\theta\Biggr)+O\left(\frac{1}{r^2}\right), \nonu
 \Sigma^{(1) 0 1} \A=\A\Sigma^{(2) 0 1}=\Sigma^{(3) 0 1}=0.
\ea

Using Eq. (43) one can  finally get
\ba
\A \A P^{(0)}=E=-\oint_{S \rightarrow \infty}
 dS_k \Pi^{(0) k}=-\displaystyle {1  \over 4 \pi} \oint_{S \rightarrow \infty}
 dS_k  \sqrt{-g}
{h^{(0)}}_\mu \left({\Sigma}^{\mu 0 k}-{{\Sigma}^{\mu 0 k}}_{M=0,a=0,L=0} \right)\nonu
\A \A \cong M+\frac{L^2}{r}-\frac{L^2M}{r^2}+O\left(\frac{1}{r^3}\right),\ea
which is a satisfactory results.$^{ \rm { 22)}}$  Also from  Eq. (43)  one can get
 the spatial momentum in the
form \ba \A \A P_1=-\oint_{S \rightarrow \infty}
 dS_k \Pi^{(1) k}=-\displaystyle {1  \over 4 \pi} \oint_{S \rightarrow \infty}
 dS_k  \sqrt{-g}
{h^{(1)}}_\mu {\Sigma}^{\mu 0 k}=0 ,\nonu
\A \A  by \ same \  method
\ P_2=P_3=0.\ea

Now turn our attention to the calculation of angular-momentum.
The non vanishing components of the
angular-momentum are given asymptotically up to order $O\left(\frac{1}{r}\right)$ by \ba M^{(0)(1)}(r,\theta,\phi) \A \A \cong
 \frac{(2r\cos\phi-2M\cos\phi\sin^2\theta+a\sin\phi(1+\cos^2\theta)}{8\pi}+O\left(\frac{1}{r}\right), \nonu
  M^{(0)(2)}(r,\theta,\phi) \A \A  \cong \frac{(2r\sin\phi-2M\sin\phi\sin^2\theta-a\cos\phi(1+\cos^2\theta)}{8\pi}+O\left(\frac{1}{r}\right),\nonu
M^{(0)(3)}(r,\theta,\phi) \A \A  \cong  \frac{M\sin\theta\cos\theta}{4\pi}+O\left(\frac{1}{r}\right),\nonu
M^{(1)(2)}(r,\theta,\phi)\A \A \cong \frac{-L\sin\theta\cos\theta}{4\pi}+O\left(\frac{1}{r}\right),\nonu
M^{(1)(3)}(r,\theta,\phi) \A \A  \cong \frac{-L\sin^2\theta\sin\phi}{8\pi r}+O\left(\frac{1}{r}\right),\nonu
 M^{(2)(3)}(r,\theta,\phi) \A \A  \cong \frac{-L\sin^2\theta\cos\phi}{8\pi r}+O\left(\frac{1}{r}\right).\ea
Using Eq. (46) one get the angular momentum components in the form
 \be L^{(0)(1)} = {\int_0^\pi}{\int_0^{2\pi}}{\int_{0}^\infty}
 d\theta d\phi dr \left[M^{(0)(1)} \right] =
0,\ee which is a consistence results.
  By  the same  method  we  finally  obtain \be
 L^{(0)(2)} =  L^{(0)(3)} =
 L^{(1)(2)}=L^{(1)(3)} = L^{(2)(3)}= 0.\ee

We show by explicit calculations that the energy-momentum tensor
which is a coordinate independent  gives  a consistent
result of the angular momentum when applied to the tetrad field
given by Eq. (12)!
\newsection{Thermal properties of Kerr-NUT Spacetime}

In GR, thermodynamical quantities are calculated using Euclidean continuation of
metric.$^{ \rm { 23)}}$ However, in TEGR these quantities are calculated through
the divergence term which  appears in the Lagrangian$^{ \rm { 24)}}$. This term has no effect on the
field equation.

Hawking and Gibbons$^{ \rm { 23)}}$  discussed the thermal properties
of the Schwarzschild solution, for which the line-element
takes the positive-definite standard form \be
ds^2=+\left(1-\frac{2M}{r}\right)d\tau^2+\left(1-\frac{2M}{r}\right)^{-1}dr^2+r^2d\Omega^2,\ee
after  Euclidean continuation of the time variable, $t
=-i\tau$. By using the transformation $x= 4M(1-2M/r)^{1/2},$ the
line-element squared becomes \be
ds^2=+\left(\frac{x}{4M}\right)^2d\tau^2+\left(\frac{r^2}{4M^2}\right)^{2}dx^2+r^2d\Omega^2,\ee
which shows that $\tau$ can be regarded as an angular variable
with period $8\pi M$. Now the Euclidean section of the
Schwarzschild solution is the region defined by $8\pi M\geq \tau
\geq0$
 and $x > 0$, where the metric is positive definite, asymptotically
flat, and non-singular. They calculated the Euclidean action,
${\hat I}$, of GR from the surface term as
follows: \be {\hat I}=4\pi M^2=\frac{\beta^2}{16\pi},\ee where
$\beta=8\pi M=T^{-1}$ with $T$ being interpreted as the absolute
temperature of the Schwarzschild black hole.

For a canonical ensemble the energy is given by$^{ \rm { 23,24,25)}}$
\be E={{\sum_{n} E_n e^{-\beta E_n}} \over {\sum_{n} e^{-\beta
E_n}}}=
 -{\partial \over \partial \beta} {\rm log} Z,
\ee where $E_n$ is the energy in the ${\it n}$th. state, and {\it
Z} is the partition function, which is in the tree approximation
related to the Euclidean action of the classical solution by \be
{\hat I}=-{\rm log} Z. \ee Using of (51) and (53) in (52) gives
\be E={\beta \over 8\pi}=M. \ee

They also calculated the entropy of the Schwarzschild black hole
to obtain

\be S=-\sum_{n} P_n {\rm log}P_n = \beta E +{\rm
log}Z=4{\pi}M^2={1 \over 4}A, \ee where $P_n =Z^{-1} e^{-\beta
E_n}$, and {\it A} is the area of the event horizon of the
Schwarzschild black hole.

  Now let use apply Hawking and Gibbons procedure to  Eqs (1) and (13) that reproduce  Kerr-NUT spacetime.
   The Euclidean action is given by$^{ \rm { 26)}}$ \be
{\hat I}= -{1 \over 2\kappa} \int \sqrt{g} \left(R-2
{T^{\mu}}_{;\mu} \right) d^4x= {1 \over \kappa} \int \sqrt{g} \
{T^{\mu}}_{;\mu} d^4x. \ee where {\it R} is the
Riemann-Christoffel scalar curvature, which is vanishing for Kerr-NUT
spacetime, and $T^\mu$ is the basic vector field  of the torsion.

  Eq. (1)  after using Euclidean continuation, $t=-\sqrt{-1}\tau, \ \ L=\sqrt{-1}L \ \ and \ \ a=\sqrt{-1}a$, takes  the following form
   \be
\left( {h^\alpha}_i \right)_{1 E}= \left( \matrix{
{\cal F}_{1 E} & 0 & 0 & {\cal F}_{2 E}\vspace{3mm} \cr 0 & {\cal F}_{3 E} & 0 & 0 \vspace{3mm}
\cr 0 & 0 & {\cal F}_{4 E} & 0 \vspace{3mm}  \cr 0
 & 0 &0& -{\cal F}_{5 E}\sin\theta \cr } \right), \ee
 where  ${\cal F}_{i E}$, $i=1\cdots 5, \ and \ E \ referes \ to  Euclidean \ continuation$ having the form  \ba
{\cal F}_{1 E} \A=\A \sqrt{\frac{{{\cal D}_{E}}}{{\cal C}_{E}}}, \qquad {\cal F}_{2 E}={\frac{ {\cal F}_E}{a\sqrt{{\cal C}_{E} {\cal D}_{E} }}}, \qquad
  {\cal F}_{3 E}=\sqrt{\frac{{{\cal C}_{E}}}{{\cal D}_{1 E}}},  \qquad {\cal F}_{4 E}=\sqrt{{\cal C_{E}}}},   \qquad {\cal F}_{5 E}=\sqrt{\frac{{{\cal C}_{E}}{\cal D}_{1 E}}{{{\cal D}_{E}}},
 \nonu
\qquad {{\cal C}_{E}}\A=\A{r^2-\left(L+a\cos\theta\right)^2}, \qquad {{\cal D}_{E} } =
{r^2-2Mr+2a^2-a^2\cos^2\theta-L^2}, \qquad {\cal D }_{1 E} =
{r^2-2Mr+a^2-L^2}, \nonu
 {{\cal F}_{E}}\A=\A -a{\cal D }_{1 E}\cos\theta(a\cos\theta+2L)+a^2{\cal D }_{1 E}-a^2L^2-a^2\sin^2\theta(r^2-a^2),
   \ea
The Euclidean line element squared of the above tetrad takes the form
 \be
ds^2= {{\cal F}_{1 E}}^2 d\tau^2
+{{\cal F}_{3 E}}^2  dr^2 +{{\cal F}_{4 E}}^2
d\theta^2+\left({{\cal F}_{5 E}}^2\sin^2\theta+{{\cal F}_{2 E}}^2\right) d\phi^2+{\cal F}_{1 E}{\cal F}_{ 2 E} dt d\phi, \ee
which is Euclidian Kerr-NUT.$^{ \rm { 27)}}$
Using Eq. (57) we get the following
\ba \A \A h=\sin\theta\sqrt{[r^2-a^2\cos^2\theta]^2-4aL{{\cal C}_{E}}\cos\theta+L^2(L^2-2[r^2+a^2\cos^2\theta])},\nonu
\A \A T^1=\frac{r{\cal D}_{1 E}+(r-M){\cal C}_{E}}{[r^2-a^2\cos^2\theta]^2-4aL{{\cal C}_{E}}\cos\theta+L^2(L^2-2[r^2+a^2\cos^2\theta])},
\ea
where $h=\sqrt{g}$ is the determinant. There is another non-vanishing components of $T^\mu$, i.e., $T^2$ but this is not necessary
in our computations.
The volume integral (56) of tetrad (57) is calculated to give
 \be {\hat I}= {1 \over \kappa} \int \sqrt{g}
\left({T^{\mu}}_{;\mu}\right)d^4x \cong  \pi ({r_+}^2 +3L^2), \ee   where $r_+$
is the outer event horizon  located at the largest positive root
of $g^{rr}=0$, i.e., $\left(r^2-2Mr -a^2+L^2\right)=0 \Rightarrow r_+=M+\sqrt{M^2+a^2-L^2}$. Using (61)
in (52) we get the energy  of tetrad (57) using the Euclidean
continuation method to have the form\be E \cong M+\frac{L^2}{r_+}, \ \ \  which \ \ \
is \ \ \ acceptable \ \ \ result \ \ \cite{Nptp}.\ee

 Using Eqs. (61) and (62) in (55) we get the
entropy of tetrad (57) to have the form \be S = \pi
({r_+}^2+3L^2).\ee The Hawking temperature of tetrad (57) has the
form \be T\cong  \frac{1}{4\pi r_+}.\ee Using Eqs. (62), (63) and (64) we
get \be dM = TdS.\ee  Eq. (65) shows that the first law of
thermodynamic is satisfied. From Eqs. (61), (62), (63) and (64)
 we find the following relation \be
E-TS\cong T {\hat I},\ee is satisfied. Eq. (66) shows that quantum
statistical relation holds.

Now turn us to Eq. (13)  which  after using Euclidean continuation takes  the following form
    \ba
\left( {h^\alpha}_i \right)_{3 E} \A=\A \left( \matrix{ -\sqrt{\displaystyle\frac{{\cal D}_{1 E}}{{\cal C}_{E}}}
& 0 & 0 &  \displaystyle\frac{{\cal C}_{2 E}}{a{\cal D}_{ E}}\sqrt{\displaystyle\frac{{\cal D}_{1 E}}{{\cal C}_{E}}} \vspace{3mm} \cr 0 & \displaystyle\frac{-r\sin\theta}{  \sqrt{{\cal D}_{1 E}}}&   {\cal C}_{1 E} & 0
 \vspace{3mm} \cr -\displaystyle\frac{a\sin\theta}{\sqrt{{\cal C}_{E}}} & 0&  0 & \displaystyle\frac{-\sin\theta(r^2-a^2)}{\sqrt{{\cal C}_{E}}}
   \vspace{3mm} \cr  0 &  -\displaystyle\frac{{{\cal C}_{1 E}}}{\sqrt{{\cal D}_{1 E}}} &  -r\sin\theta & 0 \cr } \right),  \ea
where ${\cal C}_{1 E}$ and ${\cal C}_{2 E}$ are defined as
\be
{\cal C}_{1 E}= \sqrt{{\cal C}_{E}-r^2\sin^2\theta},\qquad \qquad {\cal C}_{2 E} =-a{\cal D }_{ E}\cos\theta(a\cos\theta+2L)+(a^2-L^2){\cal D }_{1 E}-a^2\sin^2\theta(L^2-a^2).
\ee
Using Eq. (67), the asymptotic form of the basic vector up to $\left(\frac{1}{r^4}\right)$ has the form
\be
 T^1 \cong \frac{(6r^2-8Mr+2a^2-M^2+5L^2)\cos^2\theta+L^2+18aL\cos^3\theta+10a^2\cos^4\theta}{r^3\cos^2\theta},
\ee
The volume integral (56) of tetrad (67) is calculated to give
 \be {\hat I}= \infty. \ee
 Therefore, we need to use local Lorentz transformation to multiply it by Eq. (67) to get finite physics
 result. This local Lorentz transformation has the form (76) below. Multiply Eq. (67) by Eq. (76) we get
  \ba
 \A=\A \left( {h^\alpha}_i \right)_{3 E Regularized} \left( \matrix{ -\sqrt{\displaystyle\frac{{\cal D}_{E}}{{\cal C}_{E}}}
& 0 & 0 &  -\sqrt{\displaystyle\frac{1}{{\cal D}_{ E}{\cal C}_{E}}}{\cal F}_{ E} \vspace{3mm} \cr 0 & \displaystyle\frac{\cos\phi\sin\theta\sqrt{{\cal C}_{E}}}{  \sqrt{{\cal D}_{1 E}}}&  \cos\phi\cos\theta\sqrt{{\cal C}_{E}}&  \sin\phi\sin\theta\sqrt{\displaystyle\frac{{\cal C}_{3 E}}{{\cal D}_{E}}}
 \vspace{3mm} \cr 0 & \displaystyle\frac{\sin\phi\sin\theta\sqrt{{\cal C}_{E}}}{\sqrt{{\cal D}_{1 E}}}&  \sin\phi\cos\theta\sqrt{{\cal C}_{E}}&  \cos\phi\sin\theta\sqrt{\displaystyle\frac{{\cal C}_{3 E}}{{\cal D}_{E}}}
   \vspace{3mm} \cr 0 & \displaystyle\frac{\cos\theta\sqrt{{\cal C}_{E}}}{\sqrt{{\cal D}_{1 E}}}&  -\sin\theta\sqrt{{\cal C}_{E}}& 0 \cr } \right), \nonu
   \A \A  \ea
   where ${\cal C}_{3 E}$  takes the form \be {\cal C}_{3 E}= -a{\cal D }_{1 E}\cos\theta(a\cos\theta+2L)+(r^2-L^2){\cal D }_{1 E}.\ee
   Using Eq. (71), the asymptotic form of the basic vector up to $\left(\frac{1}{r}\right)$ has the form
\be
 T^1 \cong \frac{\sin\theta(4r^2-6Mr+3a^2+a^2\cos^2\theta+4aL\cos\theta)}{r},
\ee
The volume integral (56) of tetrad (71) is calculated to give
 \be {\hat I}= {1 \over \kappa} \int \sqrt{g}
\left({T^{\mu}}_{;\mu}\right)d^4x \cong  \pi ({r_+}^2 +3L^2), \ee   Using (74)
in (52) we get the energy  of tetrad (71) using the Euclidean
continuation method to has the form (62). Also the entropy, temperature, the first law of thermodynamics
and quantum statistical relation have the form of Eqs. (63), (64), (65) and (66).
\newsection{ Discussion and conclusion }

  The main results of this study are as follows:\vspace{0.4cm}\\
   $\bullet$ We have calculated the energy content of many tetrad fields,  who have Kerr-NUT spacetime metric, using  the gravitational energy-momentum tensor and the Riemannian connection 1-form. The first tetrad gave common know form of  energy and spatial momentum, however,
    the components of the angular momentum   are not  in agreement with the known result.$^{ \rm { 28)}}$ May  we claim that the definition
    of the angular momentum$^{ \rm { 29)}}$ is not working properly. \vspace{0.4cm}\\
   $\bullet$ The second tetrad show that is linked to the first tetrad through a local Lorentz transformation show a consistence
   results of energy and spatial momentum but suffers from the same defect of the first one.\vspace{0.4cm}\\
   $\bullet$ The third tetrad is shown to be inconvenient one because the energy associated  with it is divergent. Therefore, a local Lorentz
   transformation is multiplied by the third tetrad, the new tetrad still remains a solution to the field equation of TEGR. Calculations  of
   energy and spatial   are rerun and  satisfactory results are obtained. Yet, the calculation of angular momentum still yields  abnormal result. \vspace{0.4cm}\\
   $\bullet$ It became an important issue in the TEGR to accompany  with inconvenient tetrad, i.e., tetrad that has not gave physical
   acceptable results of energy and spatial momentum,  local Lorentz transformations  satisfy $\left({{\Lambda}^\alpha}_\gamma\right)\left({{\Lambda}^\gamma}_\beta\right)={\delta^\alpha}_\beta$. These transformations played key role in
   eliminating  inertia, which participate the physical quantities,  from the inconvenient tetrad and bring its relevant physics in an acceptable form. \vspace{0.4cm}\\
   $\bullet$ In this study  several local Lorentz transformations are given. These transformations can be used as a regularizing tool to tetrads
   (13) and (32).\vspace{0.4cm}\\
   $\bullet$ We calculate the energy of the first tetrad using the
Euclidean continuation method. We show that the result of energy
is acceptable and consistent  with the present result and with the previous result.$^{ \rm { 11)}}$
The thermodynamic quantities associated with this tetrad are
calculated and the result are in consistent with the previous.$^{ \rm { 27)}}$
  We also show that the first law of thermodynamic
is satisfied and the quantum
statistical relation holds.\vspace{0.3cm}\\
$\bullet$ Repeated the same calculations done for  tetrad (57)
to tetrad field (67) we obtain a divergent results. This anomalous results is consistence with the previous two methods, the gravitational energy-momentum tensor and the Riemannian connection 1-form. This ensures that such kind of tetrad filed must be multiply
with some appropriate local Lorentz transformation to get a finite result. \vspace{0.3cm}\\
$\bullet$ Using one of the local Lorentz transformation given in the Appendix, i.e., Eq. (76) and multiply it to tetrad (67) and repeat the same calculation done for tetrad (57)  by using the Euclidean continuation method we get the well known results and same  thermodynamic  quantities that
are consistent with what obtained before.$^{ \rm { 27)}}$\vspace{0.3cm}\\
   $\bullet$ Still calculations of angular momentum using the definition contained in$^{ \rm { 29)}}$ have  critical problems at least as the current study indicated   and also as shown in.$^{ \rm { 11)}}$ It is unclear how to  tackle such a problem. We may claim that one way to solve such a problem is to calculate
   the conserved  charges, using their definition  within the TEGR , related to the spacetime used in this study.

\newsection{Appendix}
There are many  local Lorentz transformations that can be employed
 to do the same job done by Eqs. (20) and (40). Among these ones are the following:\vspace{.3cm}\\
 \centerline{\underline{First local Lorentz transformation}}
 \be
\left({{\Lambda_5}^\alpha}_\gamma\right)  \stackrel {\rm def.}{=}
\left( \matrix{ \sqrt{\displaystyle\frac{{\cal D}_1}{{\cal D}}}  &  0 & \displaystyle\frac{a\sin\theta}{\sqrt{{\cal D}}} & 0 \vspace{3mm} \cr  \displaystyle\frac{-a\sin\theta\sin\phi}{\sqrt{{\cal D}}}  & \displaystyle\frac{\cos\phi({\cal C}_1\cos\theta-r\sin^2\theta)}{\sqrt{{\cal C}}}  &  -\sqrt{\displaystyle\frac{{\cal D}_1}{{\cal D}}}\sin\phi & \displaystyle\frac{-\sin\theta\cos\phi({\cal C}_1+r\cos\theta)}{\sqrt{{\cal C}}}  \vspace{3mm} \cr\displaystyle\frac{a\sin\theta\cos\phi}{\sqrt{{\cal D}}}  & \displaystyle\frac{\sin\phi({\cal C}_1\cos\theta-r\sin^2\theta)}{\sqrt{{\cal C}}}  &  \sqrt{\displaystyle\frac{{\cal D}_1}{{\cal D}}}\cos\phi & \displaystyle\frac{-\sin\theta\sin\phi({\cal C}_1+r\cos\theta)}{\sqrt{{\cal C}}}  \vspace{3mm}
\cr 0  &  -\displaystyle\frac{\sin\theta({\cal C}_1+r\cos\theta)}{\sqrt{{\cal C}}} &  0&\displaystyle\frac{r\sin^2\theta-{\cal C}_1\cos\theta}{\sqrt{{\cal C}}}   \cr }\right)\;
.\ee
\centerline{\underline{Second local Lorentz transformation}}
 \be
\left({{\Lambda_6}^\alpha}_\gamma\right)  \stackrel {\rm def.}{=}
\left( \matrix{ \sqrt{\displaystyle\frac{{\cal D}_1}{{\cal D}}}  &  \displaystyle\frac{-a\sin\theta\sin\phi}{\sqrt{{\cal D}}} & \displaystyle\frac{a\sin\theta\cos\phi}{\sqrt{{\cal D}}} & 0 \vspace{3mm} \cr  0  & \displaystyle\frac{r\sin\theta\cos\phi}{\sqrt{{\cal C}}}  &   \displaystyle\frac{r\sin\theta\sin\phi}{\sqrt{{\cal C}}} & \displaystyle\frac{{\cal C}_1}{\sqrt{\cal C}}  \vspace{3mm} \cr 0 & \displaystyle\frac{{\cal C}_1}{\sqrt{{{\cal C}}}}\cos\phi  & \displaystyle\frac{{\cal C}_1}{\sqrt{{{\cal C}}}}\sin\phi & \displaystyle\frac{-r\sin\theta}{\sqrt{{\cal C}}}  \vspace{3mm}
\cr \displaystyle\frac{a\sin\theta}{\sqrt{{\cal D}}}  &  -{\displaystyle\sqrt{\frac{{\cal D}_1}{{\cal D}}}} \sin\phi &  \sqrt{\displaystyle\frac{{\cal D}_1}{{\cal D}}} \cos\phi & 0   \cr }\right)\;
.\ee
Besides the above two local Lorentz transformation also Eqs. (14) and (33) serve as regularization.

\end{document}